\def\year{2018}\relax
\documentclass[letterpaper]{article} 
\usepackage{aaai19}  
\usepackage{times}  
\usepackage{helvet}  
\usepackage{courier}  
\usepackage{url}  
\usepackage{graphicx}  
\usepackage{booktabs} 
\usepackage{wrapfig}
\usepackage{float}
\usepackage{subcaption}
\usepackage{multirow}
\usepackage{amsmath}
\usepackage[colorinlistoftodos,prependcaption,textsize=tiny]{todonotes}
\usepackage[ruled]{algorithm2e} 

\newcommand{\citet}[1]{\citeauthor{#1}\shortcite{#1}}

\begin{document}
%
\title{Deep Dive into Anonymity: A Large Scale Analysis of Quora Questions}

\author{Binny Mathew\textsuperscript{$1$}\thanks{contributed equally to this paper.}, Ritam Dutt\textsuperscript{$1*$}, Suman Kalyan Maity\textsuperscript{$2$}, Pawan Goyal\textsuperscript{$1$}, Animesh Mukherjee\textsuperscript{$1$}\\
	\textsuperscript{$1$} Indian Institute of Technology, Kharagpur\\
	\textsuperscript{$2$} Northwestern University\\	
	{\tt \small binnymathew@iitkgp.ac.in, ritam.dutt@gmail.com , suman.maity@kellogg.northwestern.edu} \\
	{\tt \small pawang.iitk@gmail.com, animeshm@gmail.com}%
}

\maketitle
\begin{abstract}
Anonymity forms an integral and important part of our digital life. It enables us to express our true selves without the fear of judgment. In this paper, we investigate the different aspects of anonymity in the social Q\&A site Quora. The choice of Quora is motivated by the fact that this is one of the rare social Q\&A sites that allow users to explicitly post anonymous questions and such activity in this forum has become normative rather than a taboo. Through an analysis of 5.1 million questions, we observe that at a global scale \textit{almost no difference} manifests between the linguistic structure of the anonymous and the non-anonymous questions. We find that topical mixing at the global scale to be the primary reason for the absence. However, the differences start to feature once we ``deep dive'' and (topically) cluster the questions and compare the clusters that have high volumes of anonymous questions with those that have low volumes of anonymous questions. In particular, we observe that the choice to post the question as anonymous is dependent on the user's perception of anonymity and they often choose to speak about \textit{depression}, \textit{anxiety}, \textit{social ties} and \textit{personal issues} under the guise of anonymity. We further perform personality trait analysis and observe that the anonymous group of users has positive correlation with \textit{extraversion}, \textit{agreeableness}, and negative correlation with \textit{openness}.  Subsequently, to gain further insights, we build an \textit{anonymity grid} to identify the differences in the perception on anonymity of the user posting the question and the community of users answering it. We also look into the \textit{first response time} of the questions and observe that it is lowest for topics which talk about personal and sensitive issues, which hints toward a higher degree of community support and user engagement.

\end{abstract}

\section{Introduction}

\subsection{Anonymity in Q\&A sites}
Question answering sites are one of the primary sources on the Internet that attempt to meet this huge information need of the users. Q\&A sites like Yahoo! Answers, Quora, Stack Exchange are community efforts that provide answers to questions on a wide range of topics. Some of these sites like Yahoo! Answers and Quora have a unique feature that allows users to post questions anonymously which enables them to ask judgmental or controversial questions freely. Note that these are different from social media sites like Whisper and Secret because while the former allow posts to be both anonymous or non-anonymous depending on the user's preference, the latter allow posts to be strictly anonymous. Quora, in addition, has two interesting and very rich features -- (i) the questions can be organized topically that allows the users to search for relevant answers easily and the moderators to manage and route/promote questions to appropriate communities for garnering better answers and (ii) there is an underlying social network that allows user to follow other users, questions and topics. Over the years these have immensely enriched the interactions within the forum thus improving both the question and the answer quality.

\subsection{Anonymity in Quora contrasts other online forums}
Another important point to note here is that while anonymous online forums have been around for a long time, they are plagued by several issues like trolling, relatively small user population, lack of expertise etc. In contrast, Quora does not suffer from these issues. The user base is substantially larger consisting of people from all walks of life many of whom are actually experts in various domains.  This fosters a rich atmosphere where people can provide quality answers based on their expertise. Also in the Quora ecosystem, asking questions anonymously is as much acceptable to the community as asking non-anonymously; as high as 38.7\% questions are asked anonymously on Quora. Further, the acceptance of a question from the community is dependent on the responses it has received. Consequently, we observe that the mean number of responses for the anonymous and non-anonymous questions on a global scale is approximately 2.86 and 2.99 respectively. Users resort to asking questions anonymously because they feel that it has various benefits like revealing the ``truth''\footnote{https://www.quora.com/Why-did-Quora-add-the-anonymous-option-for-answers-and-questions} or asking sensitive questions like ``I am depressed. How do I hide it at school?'' without the fear of its repercussions. The practice has become normative and no longer a taboo or something to be looked down upon. Finally, the issue of trolling is also tackled well in Quora through strict regulatory policies enforced by the moderators that are normally adhered to by the community at large.

\subsection{Key contributions and observations in this paper}
Analysis of a large dataset of 5.1 million Quora questions reveals that around 38.7\% of them are posted anonymously. This observation forms the primary motivation of the current paper in which we present a detailed and extensive study characterizing the anonymous questions on Quora. We summarize our main contributions and observations in this paper as follows.

\begin{itemize}

\item We observe \textit{almost no difference} between the linguistic structure of the questions posted anonymously and those posted non-anonymously at a global scale.

\item We identify a substantial proportion of topical mixing at the global scale i.e., questions that should be apparently posted as anonymous are posted as non-anonymous and vice versa. Oftentimes, such a behavior can be attributed to the user's perspective~\cite{correa2015many} of what should be posted as anonymous and what not.

\item As a next step, we ``deep dive'' into the data, whereby, we (topically) cluster the questions and individually inspect the clusters containing large volumes of anonymous questions and compare them with those that predominantly contain non-anonymous questions. Nuanced linguistic differences start becoming apparent revealing that users extensively query about \textit{depression}, \textit{anxiety}, \textit{social ties} and \textit{personal issues} under the guise of anonymity.

\item Personality trait analysis further reveals that user groups posting anonymous questions show positive correlation with traits like \textit{extraversion} and \textit{agreeableness}. In contrast, they seem to bear negative correlation with the trait \textit{openness}.

\item We establish the concept of \textit{anonymity grid} to identify the differences in the perception of anonymity between the user posting the question and the community answering the same. While personal questions related to {\sl sex} and {\sl love life} are both anonymously asked and answered, questions related to {\sl job} and {\sl relationship advice} are asked anonymously but answered non-anonymously and questions related to incensed topics like {\sl god}, {\sl religion} and {\sl politics} are asked non-anonymously but answered anonymously.

\item We also leverage the concept of First Response Time (FRT) to quantify the time taken for a question to receive its first response; a lower FRT corresponds to a quicker response. We note that although there is no significant difference in the median FRT for anonymous and non-anonymous questions on a global scale, it becomes more pronounced once we topically cluster them. Specifically, the FRT for topics having predominately anonymous questions is sufficiently lower than those with non-anonymous questions. In fact, we observe the lowest FRT for extremely personal topics such as \textit{anxiety, depression, sexual violence etc}. This demonstrates a high level of community support enhancing user engagement which seem to have evolved into an intrinsic and integral characteristic of Quora, hitherto unreported.

\end{itemize}

\section{Related work}

There have been several research in the past to study anonymity in the online and offline settings. In this section, we present a comprehensive review of the different aspects of anonymity research.

\subsection{Effects of anonymity}

Anonymity can lift inhibition which can lead to unusual acts of kindness or generosity, or it can lead to misbehavior, such as harsh or rude language and acts that are illegal or harmful~\cite{suler2004online}. Christopherson~\shortcite{christopherson2007positive} discusses the positive and the negative effects of anonymity in detail.

There are several studies focusing on the negative aspects of anonymity, such as cyberbullying~\cite{huang2010analysis}, aggressive behaviour~\cite{zimbardo1969human}, encouraging suicidal individuals to follow through with their threats~\cite{mann1981baiting}, hate sites~\cite{chau2007mining}, and many more.

Another set of studies detail the positive aspects of anonymity mainly involving self-disclosure and degree of intimacy. In some research, the absence of interviewer has been found to increase the duration of self-disclosure for participants who were presented with intimate questions~\cite{jourard1970experimenter}. Privacy has been shown to have positive effect on human well-being in~\cite{werner1992transactional}. De et al.~\shortcite{de2014mental} find that anonymity helps users to discuss topics which are considered to be stigma in the real world. People regularly use the protection of anonymity to reduce the social risks and to create different persona online than they exhibit offline~\cite{yurchisin2005exploration,bargh2002can}. In Yurchisin et al.~\shortcite{yurchisin2005exploration}, the authors find that the users use online dating services in identity exploration and recreation process. Anonymity also helps in the protection of informants (e.g., whistle-blowers or news sources).

\subsection{Anonymity in social media}

Several social media sites allow options to post content anonymously. Facebook provides `confession pages' while Yahoo! Answers and Quora allow users to post questions and answers as anonymous. There are several mobile services that allow anonymous sharing like Whisper and Secret. 

In Correa et al.~\shortcite{correa2015many}, authors study the sensitivity and types of content posted on Whisper and Twitter. They find that the anonymity of `whispers' is not binary, implying that the same question text exhibit different levels of anonymity, depending on the user's perception. In addition, they find significant linguistic differences between whispers and tweets. In Birnholtz~\shortcite{birnholtz2015weird}, the authors study the questions that are posted on Facebook confession boards (FCBs). They find that users ask about taboo and stigmatized topics and tend to receive relevant responses that are potentially useful and contain less negativity. In Bernstein et al.~\shortcite{bernstein20114chan}, the authors study online ephemerality and anonymity on /b/ board in 4chan. They find that over 90\% of the posts in /b/ board are made anonymously. In De Choudhury and De~\shortcite{de2014mental}, the authors study the impact of anonymity on sharing of information related to stigmatic health topics. In~\citet{andalibi2016understanding}, the authors investigate self-disclosures by sexual abuse survivors and exploring the relationships between anonymity, disclosure, and support seeking by analyzing posts  from both throwaway (providing more anonymity) and identified accounts on abuse-related subreddits. In ~\citet{forte2017privacy}, the authors examine privacy practices and concerns among contributors to open collaboration projects and show that privacy concerns pervade open collaboration projects; risks are perceived both by individuals who  occupy central leadership roles in projects. In ~\citet{ma2016anonymity}, the authors conducted an online experiment to study the relationship between content intimacy and willingness to self-disclose in social media, and how identification and audience type moderate that relationship.

\subsection{Reasons for anonymity}

Internet users tend to care a lot about their privacy. In Kang, Brown, and Kiesler~\shortcite{Kang:2013:WPS:2470654.2481368}, the authors interviewed people who use anonymous communication application and found that users predominantly disclose personal information or emotions in the guise of anonymity.

For instance, Internet users cite the following reasons on the Quora Q\&A platform for posting anonymous content.

\begin{itemize}
\item To avoid other users from forming a bias based on my nationality, age, creed, caste, sexuality, sex etc. while reading my answers\footnote{https://www.quora.com/What-do-people-use-the-anonymity-feature-for-on-Quora}.
\item Users may not want to reveal information about themselves unnecessarily\footnote{https://www.quora.com/Why-do-people-post-content-to-Quora-as-anonymous-user/}.
\item Many users tend to rather share content anonymously than remain silent in fear of being judgmental or controversial.
\end{itemize}

\subsection{Research on Quora}
In~\cite{wang2013wisdom}, the authors perform a detailed analysis of Quora. They show that heterogeneity in the user and question graphs are significant contributors to the quality of Quora's knowledge base. They show that the user-topic follow graph generates user interest in browsing and answering general questions, while the related question graph helps concentrate user attention on the most relevant topics. Finally, the user-to-user social network attracts views, and leverages social ties to encourage votes and additional high quality answers. Maity et al.~\shortcite{maity2015analysis} study the dynamics of temporal growth of topics in Quora and propose a regression model to predict the popularity of a topic. In Patil and Lee~\shortcite{patil2016detecting}, the authors analyze the behavior of experts and non-experts in five popular topics and extract several features to develop a statistical model which automatically detect experts. In Pan et al.~\shortcite{pan2013answer}, the authors study how the non-Q\&A social activities of Quorans can be used to gain insight into their answering behavior. In Maity et al.~\shortcite{maity2017language}, the authors find that the use of language while writing the question text can be a very effective means to characterize answerability and it can help in predicting early if a question would eventually be answered.

There have been very few studies on anonymity in Quora. In Paskude and Lewkowicz~\shortcite{paskuda2015anonymous}, the authors analyze the anonymous and non-anonymous answers in the health category of Quora and found that anonymous answers and social appreciation correlates with the answer's length. Our work is different from this as we primarily focus on questions instead of answers. 

\subsection{Present work}
Ours is the first large-scale measurement study to characterize anonymity in Quora. We show for the first time that differences that do not feature at the global-scale become apparent when one deep dives into more microscopic analysis and investigate the (topical) clusters that are dense in anonymous questions. We build \textit{Anonymity Grid}, that takes into consideration both the users and communities perspective on anonymity and provide interesting insights. We also use the concept of \textit{first response time} and observe that the anonymous questions receive faster responses.

\section{Dataset description}
We obtain the Quora dataset from the authors of~\cite{maity2015analysis} and build on it to have a massive set of 5,160,765 questions and 488,122 topics (along with the topic logs). The questions contain information regarding the date of creation, the user\footnote{We did not remove the users whose accounts were deleted. On deletion, the username is simply replaced with ``user". Note that the deletion of a user account does not make it anonymous. Instead a placeholder username is assigned to the question and therefore we continue to treat it as a non-anonymous question. For more details please visit: https://www.quora.com/What-happens-when-I-deactivate-or-delete-my-Quora-account} who has posted the questions and also answered them, the topics assigned to the question etc. Table~\ref{tab:dataset1_properties} lists some of the numbers related to the dataset. 

\begin{table}[htb]
	\small
	{\begin{tabular}{| p{6.50cm} | p{1.25cm }|} 
	\hline
	Dataset properties & Number \\
	\hline
	\# Questions extracted & 5,160,765 \\
	\# Anonymous questions & 1,997,474 \\
	\# Non-anonymous questions &
	3,163,291 \\
	\#Answers of anonymous questions & 5,727,278\\
	\#Answers of non-anonymous questions & 9,464,554 \\
	\#Anonymous answers of anonymous questions & 589,569\\
	\#Anonymous answers of non-anonymous questions & 727,088\\
	\#Deleted answers of anonymous questions & 570,285\\
	\#Deleted answers of non-anonymous questions & 873,365\\

	\hline
\end{tabular}}
\caption{Properties of the dataset.}
~\label{tab:dataset1_properties}
 \end{table}

  

\section{Rise of anonymity in Quora}

\begin{figure}[htb]
	\centering
	\includegraphics[width=.45\textwidth]{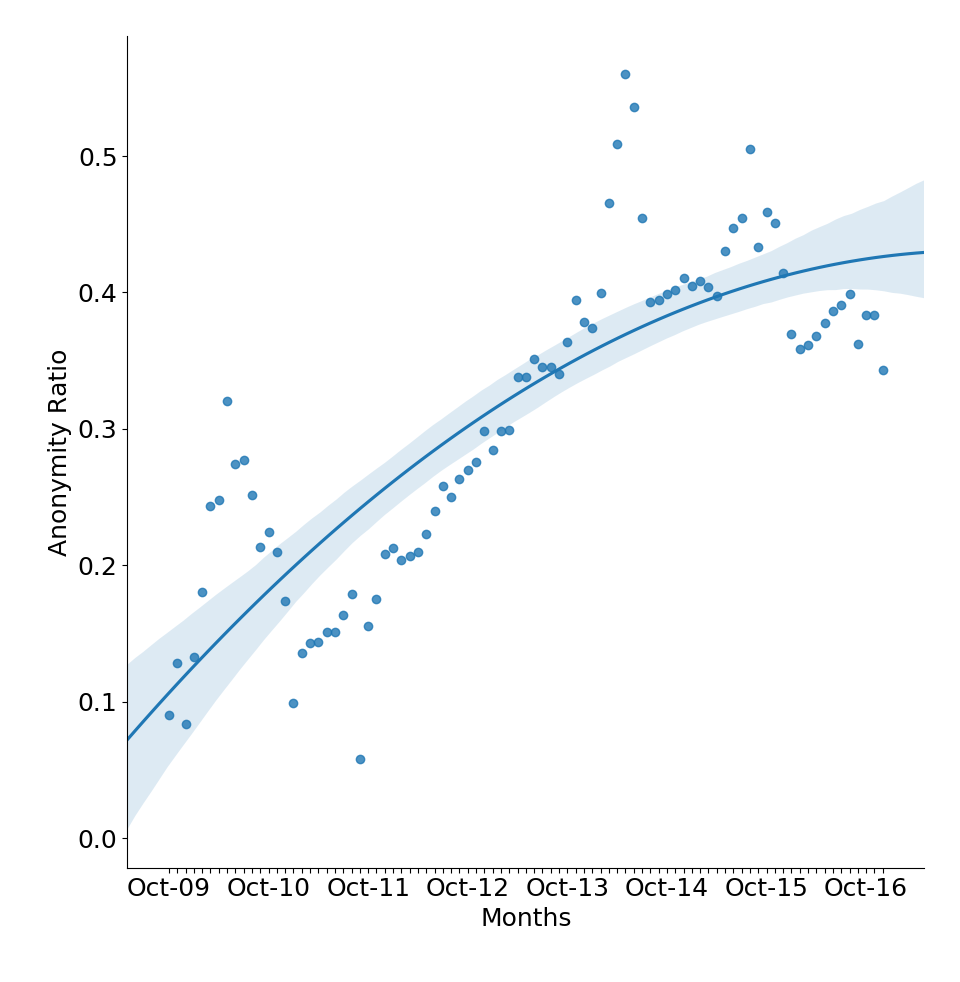} 
	\caption{Ratio of the anonymous questions to the total number of questions over time. The plotted line is obtained by fitting a quadratic curve through the datapoints while the shaded area represents the 95\% confidence interval. }
	~\label{fig:anonymity_timeline}    
\end{figure}

In our initial experiments, we measure the prevalence of the usage of the anonymity feature by the Quora users for asking questions. We find from our dataset that approximately 38.7\% of the questions have been posted as anonymous. In Figure~\ref{fig:anonymity_timeline}, we plot the ratio of the questions posted as anonymous to the total number of questions posted over time. Clearly, there is sharp increase in this ratio over time. We looked into some of the anonymity peaks to find the reasons. For example, we looked into the topics responsible for the anonymity peak on May-2010 and found majority of the anonymous questions were posted in topics related to \textit{Startups and Entrepreneurship}. Similarly, for the anonymity peak in May-2014, the majority of the anonymous questions were asked in topics related to \textit{Love, Dating, Sex} and \textit{Friendship}. 
Furthermore we note that the mean response for anonymous and non-anonymous questions are 2.86 and 2.99 respectively. However, the fraction of anonymous answers garnered by the anonymous questions is 10.29\% as opposed to 7.68\% by the non-anonymous questions. 

\section{Analysis at the global scale}

As an initial step, we segregate the questions into two groups -- those that were tagged as anonymous (henceforth 'Anon') and those that were tagged as non-anonymous (henceforth 'Known') and perform analysis on the two groups separately.

\begin{figure}[h]
	\centering	
	\begin{subfigure}[b]{0.45\linewidth}
		\includegraphics[width=\textwidth]{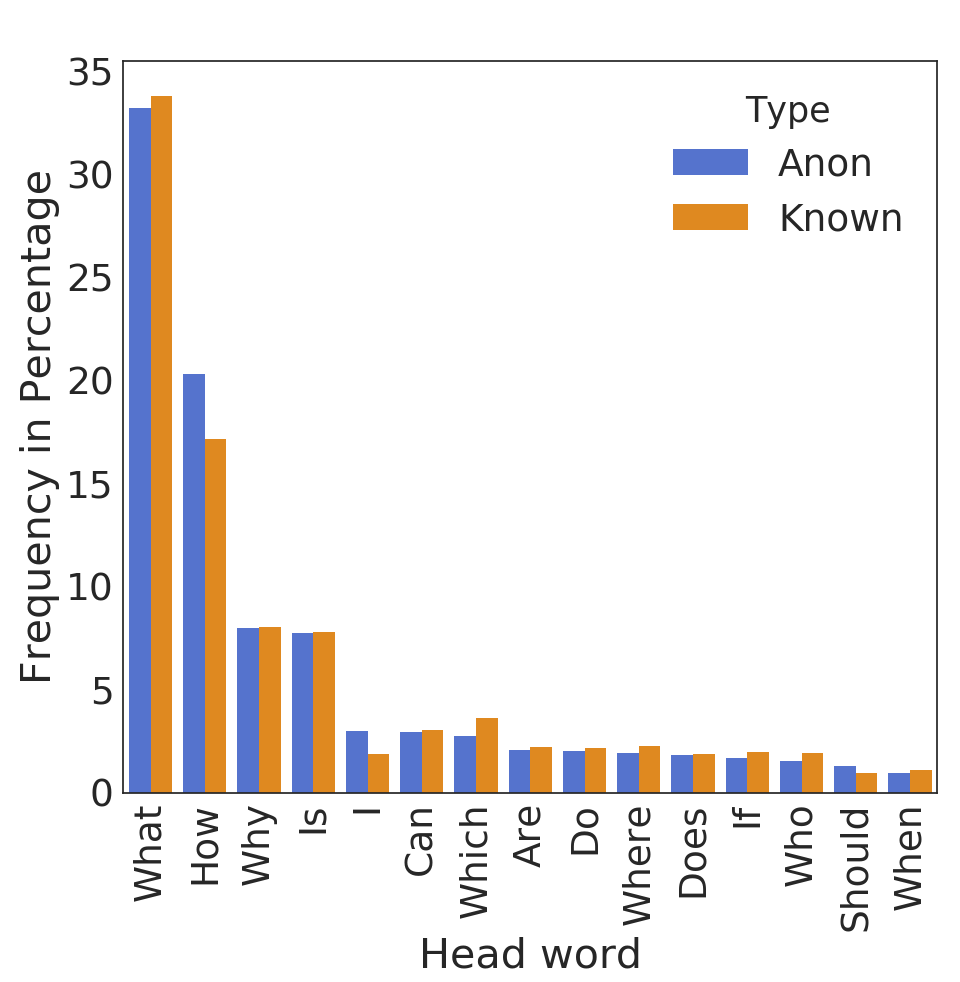}
		\caption{Global Level}
		\label{fig:question_type_global}
	\end{subfigure}
    \enskip
	\begin{subfigure}[b]{0.45\linewidth}
		\includegraphics[width=\textwidth]{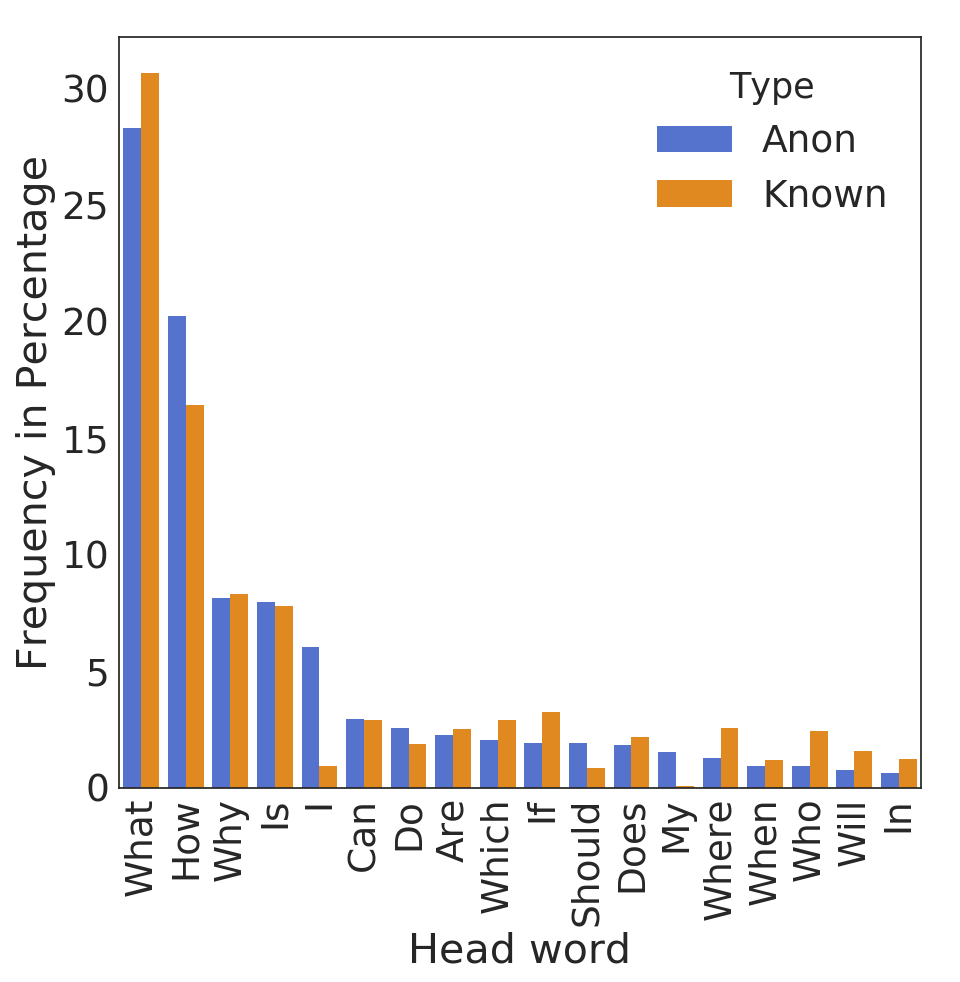}
		\caption{Topic Level}
		\label{fig:question_type_topic}
	\end{subfigure}
   \caption{Distribution of different types of head words at the global level and at the topical level.} 
\end{figure}

\noindent\textbf{Question types:}
We observe that majority of the questions posted on Quora are interrogative sentences. The head word of a question, i.e., words like ``what'', ``why'' are a key feature used in traditional question  classification~\cite{li2008classifying,huang2008question}. We consider any word that occurs at the beginning of a sentence as a head word and we check their frequency of occurrence. In Figure~\ref{fig:question_type_global}, we illustrate the proportion of these head words in the two question groups -- Anon and Known. From the figure, we observe that Anon group tends to feature more ``how'' type questions than the Known group. This could indicate that the Anon questions are asking for detailed reasons. The Known group tends to feature more ``which'', ``where'' and ``what'' type questions. This indicates that Known group tends to ask for specific information that are open-ended or restricted range of possibilities, or regarding `some place'.

\noindent\textbf{Linguistic analysis:}
Previous literature on anonymity ~\cite{correa2015many,birnholtz2015weird} have noted that there are significant differences between the text posted as Anon and those posted as Known. We thus attempt to verify whether this claim holds for Quora as well. To this end, we consider the two groups of question text -- one comprising all questions tagged as Anon while the other comprising all questions tagged as Known.
For each group we perform the following linguistic analysis namely, standard POS tag analysis (i.e., the fraction of different parts-of-speech in the text),\footnote{We use the POS (Parts of Speech) tagger of SpaCy: https://spacy.io/}, sentiment analysis\footnote{We use  VADER~\cite{hutto2014vader}.} (i.e., the fraction of positive, negative and neutral sentiment words in the text) and LIWC analysis\footnote{http://liwc.wpengine.com/} (i.e., the fraction of words in different linguistic and cognitive dimensions identified by the LIWC tool) on the question text. Finally, we look for statistically significant differences between these two groups with respect to each of the above analysis. As the size of our sample is very large, standard $t$-test would be very ineffective as it will transform small differences into statistically significant differences - even when they are practically insignificant~\cite{faber2014sample}. Therefore, we use Cohen's $d$~\cite{cohen1988statistical} to find differences that are practically significant. We do not find evidence of strong differences between the two groups. These findings are in agreement with the previous research done on Quora~\cite{paskuda2015anonymous}. In Table~\ref{tab:global_characteristics} we note those fractions that are mildly significant (as per the Cohen's $d$-test).\\

\noindent\textbf{Cohen's $d$}: The Cohen's $d$ test is primarily used to measure the effect size, which is an estimate of the strength of the relationship of two variables. Given the mean and standard deviation of two populations, denoted by $\mu_{1}$ and $\sigma_{1}$ and $\mu_{2}$ and $\sigma_{2}$ respectively, the Cohen's $d$-test is the ratio of the difference between their means to their pooled standard deviation, more succinctly represented by the following expression
\begin{center}
\large
$d = \frac{\mu_{2} - \mu_{1}}{\sqrt{\frac{\sigma_{1}^2 + \sigma_{2}^2}{2}}}$
\end{center}
A Cohen's $d$-value of magnitude 0.2 indicates small effect, 0.5 indicates medium effect while 0.8 signifies large effect.\\

\noindent\textbf{Diff(\%)}: We use the metric $Diff(\%)$ in a way similar to~\cite{correa2015many} to quantify the difference in features between the Anon and the Known Group. The Diff(\%) metric is simply the mean difference of a feature between the Anon and Known group expressed as a percentage and is represented by
\begin{center}
\large
$ Diff = \frac{Anon - Known}{Known}\times 100\%$
\end{center}


\begin{table}[htb]
\resizebox{\linewidth}{!}{\begin{tabular}{|p{3.9cm} | p{1.5cm} | p{1.0cm}|p{1.0cm}|p{1.3cm}|} 
	\hline		
	Category & Cohen's $d$ & Anon & Known& Diff. (\%) \\
	\hline
	PRON (Pronoun) &0.1413&0.0395&0.0325&21.476 \\
	\hline
	PRP\$ (Personal pronoun) &0.1413&0.0115&0.0103&11.464\\
	RBS (Adverb, Superlative) &-0.0372&0.0013&0.0018& -24.168\\
	JJS (Adjective, Superlative)&-0.0656&0.0054&	0.0068&-21.804\\
    \hline
	Negative &0.0611 &0.0442 &	0.0383& 15.349\\
	Positive &-0.0310&0.0984 &	0.1026&-4.119\\
	\hline
	Family &0.0552 &0.0019&0.0012&	55.429\\
	Sexual &0.0892 &0.0032&	0.0016&	89.284\\
	Friends &0.0648 &0.0023&	0.0014&68.024\\
	I &0.1404 &0.0265&	0.0195&	36.330\\
	We &-0.055 &0.0024 &0.0034&	-28.931\\
	\hline
	\end{tabular}}
	\caption{No significant linguistic differences between the questions of the Anon and the Known group at the global scale. The significance is measured by the Cohen's $d$ value. \\}
	~\label{tab:global_characteristics}
\end{table}

\subsection{Absence of significant linguistic differences at the global scale}

Correa et al.~\shortcite{correa2015many} acknowledges that anonymity is not a binary concept and it is perceived differently by different people. The veracity of the above claim is established upon closer examination of the Anon and Known questions in our dataset. We note instances of Known questions that talk about topics prevalently associated with some form of stigma or taboo. Some Anon questions, on the other hand, seem to be very general in nature. The question 'What building is similar to the Taj Mahal in form?' is tagged anonymously but the associated topics of the question have predominately non-anonymous questions. The reverse also holds true for the question 'How do I get rid of mental depression?' which was tagged non-anonymously. This ambiguity arises since we expect the anonymity of a question to be representative of its associated topics, which need not necessarily hold true. Table~\ref{tab:reason_anon_known} shows more examples of such `ambiguously' tagged questions. To quantify this ambiguity, we introduce the concept of anonymity ratio ($a_r$) for a topic and subsequently, the Question Anonymity Score (QAS) for a question. We then illustrate the principle of global scale topical mixing using these concepts.

\subsection{Topical mixing at global scale}
\begin{figure}[htb]
	\centering
	\includegraphics[width=.45\textwidth]{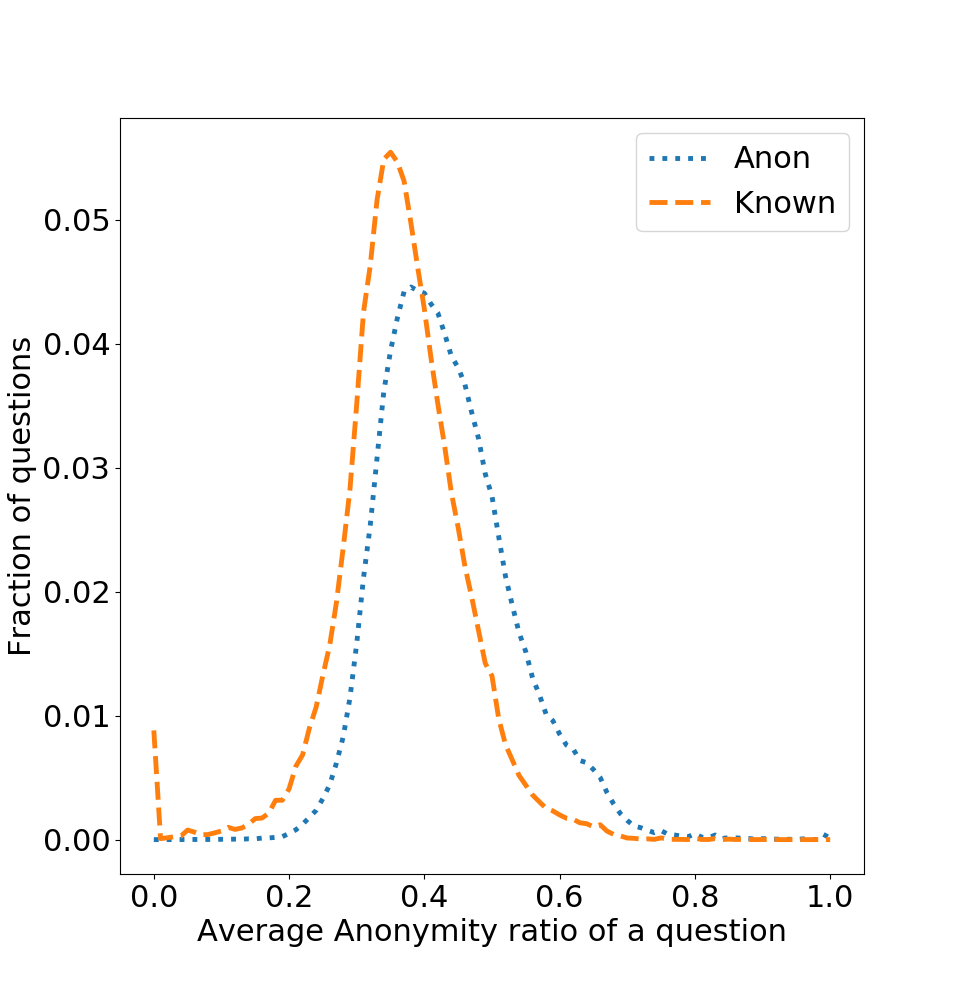} 
	\caption{Anonymity ratio of questions.}
	~\label{fig:AR_questions}    
\end{figure}

We define the \textit{anonymity ratio} ($a_r$) of a topic as the ratio of the anonymous questions to the total questions in that topic. So, each topic will have an $a_r$, with value between zero and one with one signifying that all the questions posted in this topic are anonymous. Next, we use the $a_r$ to the check the anonymity of a question. The anonymity score for a question, which we term the `Question Anonymity Score (QAS)' is defined as the mean $a_r$ of all the topics that the question is tagged with. Thus, if a question is tagged with topics that have high $a_r$, the question will also receive a high QAS. We divide the questions into two sets based on whether the question was posted as anonymous or non-anonymous, Anon and Known respectively. Figure~\ref{fig:AR_questions} shows the distribution of QAS associated with the Anon and the Known questions. As seen from the figure, there is a huge overlap between the Anon and the Known questions. We refer to this phenomenon as 'topical mixing' and reckon that this is the primary reason why we do not observe a significant difference between the Anon and the Known groups at the global scale. This forms the primary motivation for 'deep-dive' into anonymity.

\begin{table*}[htbp]
	
	\resizebox{\textwidth}{!}{\begin{tabular}{| c | c |} 
			\hline
			\textbf{Anon} &\textbf{Known} \\ \hline
			What building is similar to the taj Mahal in form?&How do I get rid of mental depression?\\ 
			How would you describe a river?&I am 17 and still single. Is it normal? \\
			what are the best google summer of code projects& I missed my period. Am I pregnant? \\
			What are the major landforms in Egypt?& How can I control my abusive nature? When I am abusive, I don't see how I am.\\
			What is the best camera ever?& I am currently facing abuse at home. I am only 17, what should I do?\\
			
			\hline
	\end{tabular}}
	\caption{Ambiguously tagged questions -- Known questions pertain to topics that correspond to some form of stigma or taboo while Anon questions are more general in nature.}
	~\label{tab:reason_anon_known}
\end{table*}

\section{Topical anonymity}
The lack of significant difference between the Known and the Anon group at the global scale further motivated us in making more granular investigations. One way to make such fine-grained investigation would be to analyze a group of questions together. We therefore attempt to group the questions in topical clusters and study their characteristics. We make use of the topics assigned to the questions for this purpose.

\subsection{Quora topic based clusters (TC)}
Topics assigned by users impose one such natural grouping on the questions. We consider only those topics which have at least 20 questions to ensure that the topic is sufficiently represented.

\begin{table}[htbp]
	\centering
	{\begin{tabular}{| c | c |} 
	\hline
	Cluster property & TC \\
	\hline 
	$\mu_{a_r}$ &0.3903\\
	$\sigma_{a_r}$ &0.1528\\
	\#Anonymous clusters &7617\\
	\#Non-anonymous clusters &6881\\
	\#Neutral clusters &36241\\
			\hline
	\end{tabular}}
	\caption{Basic statistics of the clusters obtained.}
	~\label{tab:cluster_statistics}
\end{table}

\subsection{Analysis of the topical clusters} 
The topic based clustering aims at segregating the Quora questions into three major types of clusters -- (i) those that contain predominantly anonymous questions which we call \textit{anonymous clusters} (ii) those that contain predominantly non-anonymous questions which we call \textit{non-anonymous clusters} and (iii)  those that we call \textit{neutral clusters} and are somewhere in between (i) and (ii). 

\begin{figure}[htb]
	\centering	\includegraphics[width=.50\textwidth]{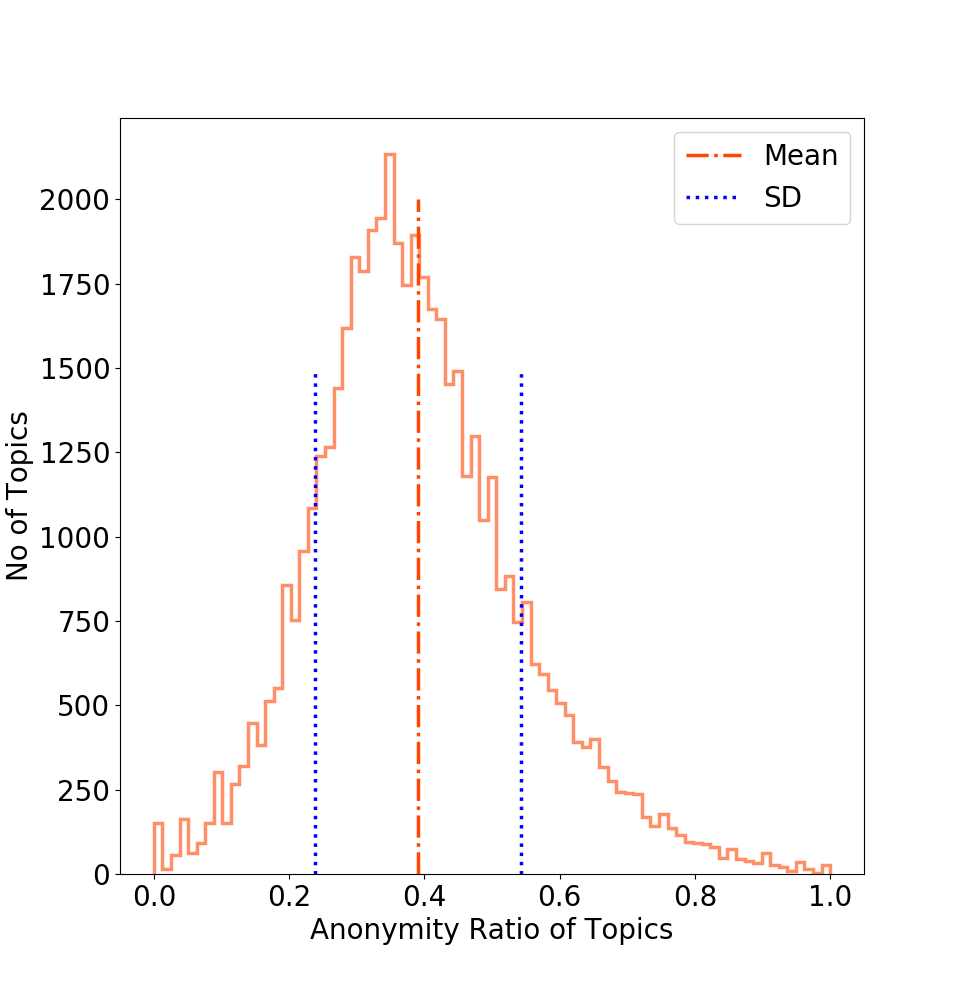} 
	\caption{Anonymity ratio ($a_r$) across the different clusters for topic based clusters.}
	~\label{fig:cluster_density}
\end{figure}

We reuse the definition of \textit{anonymity ratio} ($a_r$) of a cluster as the ratio of the anonymous questions to the total questions in that cluster. Let the mean $a_r$ across all the clusters be denoted by $\mu_{a_r}$; similarly, let the standard deviation be denoted by $\sigma_{a_r}$. We define a cluster as anonymous (Anon) if its $a_r \geq \mu_{a_r} + \sigma_{a_r}$. Similarly, we call a cluster non-anonymous (Known) if $a_r \leq \mu_{a_r} - \sigma_{a_r}$. The clusters having $a_r$ values in between these two extremes are the neutral ones.
Table~\ref{tab:cluster_statistics} lists some of the properties of the clusters generated. Figure ~\ref{fig:cluster_density} plots the distribution of $a_r$ for the clusters. 

In the following, we repeat the analysis earlier reported at the global scale for each of the Anon and the Known clusters.

\noindent\textbf{Question types}: We observe that the differences in the head word distribution becomes more prominent in topical clusters as shown in Figure~\ref{fig:question_type_topic}. Specifically questions that start with 'I' are more prominent among the Anon group than the Known ones.

\subsubsection{Structural analysis} We perform simple structural analysis on the question-texts of the Anon and Known clusters such as word distribution, question length and POS (Parts-of-Speech) analysis. 

\begin{figure}[h]
	\centering	
	\begin{subfigure}[b]{0.47\linewidth}
		\includegraphics[width=\textwidth]{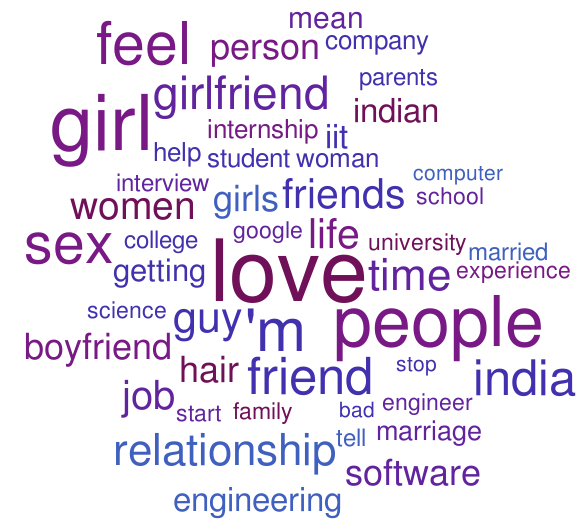}
		\caption{Words from questions in Anon clusters.}
		\label{fig:word_cloud_anon}
	\end{subfigure}
    \qquad
	\begin{subfigure}[b]{0.43\linewidth}
		\includegraphics[width=\textwidth]{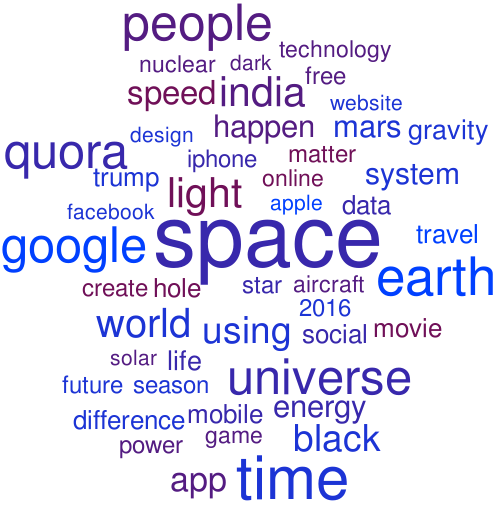}
		\caption{Words from questions in Known clusters.}
		\label{fig:word_cloud_non_anon}
	\end{subfigure}
  \caption{Word cloud of questions in Anon and Known clusters.} 
\end{figure}

\noindent\textbf{Word usage} We observe that the average word count for the Anon clusters is 15.17, while for Known clusters it is 13.81. This is due to the questions being more expressive in the Anon clusters. As we observe from the word cloud of the Anon clusters in Figure~\ref{fig:word_cloud_anon}, people tend to use anonymity for personal questions like those related to relationship, job, love,  etc., while users ask questions non-anonymously for general topics such as space, technology, Quora etc. (see Figure ~\ref{fig:word_cloud_non_anon}).

\noindent\textbf{POS analysis } To inspect any linguistic differences in the framing of question texts, we analyze the usage of POS tags in the question texts.
We compute the distribution of different POS tags (the simple POS tags follow the Universal Dependency Scheme\footnote{$http://universaldependencies.org/u/pos/$} while the finer categories correspond to the OntoNotes 5 version\footnote{$https://catalog.ldc.upenn.edu/LDC2013T19$} of the Penn Treebank tag set)
for both the Anon and the Known topics. We observe from Table~\ref{tab:POS_values} that the tags like verbs and pronouns strongly prevail in the Anon topics, while nouns seem to be in higher usage in Known topics. The questions in Anon cluster use more pronouns indicating that people use anonymity to ask more personal questions. The higher presence of VB and VBP in the Anon cluster express questions about habits and routines, thoughts and feelings~\footnote{\url{https://sites.google.com/site/partofspeechhelp/#TOC-VBP}}. The lower presence of nouns in Anon cluster is unsurprising since anonymous users would prefer not to disclose their identities by referring to persons, facilities, locations etc.

\begin{table}[htbp]
	\resizebox{\linewidth}{!}{\begin{tabular}{| p{4.5cm} | p{1.5cm} | p{1.0cm} |p{1.0cm}|p{1.5cm}|} 
			\hline
	\multicolumn{5}{|c|}{Part-of-Speech Tag (Simple)}\\
	\hline
	POS category & Cohen's $d$  & Anon & Known & Diff. (\%) \\ \hline
	NOUN    &	-0.402   & 0.276  &	0.3245 &	-14.826 \\ 
	VERB    &   0.236    & 0.220  &	0.199  &	10.549  \\
    PRONOUN &	0.569    & 0.065  & 0.029  &	124.966 \\
	
	\hline 
	\multicolumn{5}{|c|}{Part-of-Speech Tag (Detailed)}\\
	\hline 
POS category & Cohen's $d$ & Anon & Known & Diff. (\%) \\ \hline
NN	(Noun, singular)&	-0.357	&	0.187	&	0.227	&	-17.423 \\ 
VB	(Verb, base form)&	0.2	&	0.061	&	0.049	&	24.211 \\ 
VBP (Verb, non-3rd person, singular present)	&	0.226	&	0.052	&	0.038	&	35.389 \\ 
PRP\$ (Possessive Pronoun)	& 0.328	&0.02	&	0.008	& 142.843 \\ 
PRP (Personal Pronoun)	&	0.569	&	0.065	&	0.029	&	124.966 \\ 

	\hline
	\end{tabular}}
	\caption{Fraction of different POS tags across the Anon and the Known clusters. Only cases with significant difference are shown.}
    ~\label{tab:POS_values}
\end{table}

\begin{table}[h]
	\resizebox{\linewidth}{!}{\begin{tabular}{| p{3.0cm} | p{1.6cm} | p{1.2cm} |p{1.2cm} |p{1.3cm} |} 
			\hline	
	Sentiment Score & Cohen's $d$ & Anon & Known & Diff. (\%) \\ \hline
    Neutral & -0.298 & 0.826&	0.874&	-5.496\\
    Positive &  0.157&0.112&	0.091&	23.869	\\
	Negative & 0.248 & 0.061& 0.035& 75.286\\    

\hline
	\end{tabular}}
	\caption{Sentiment score of VADER across the Anon and the Known topics.}
	~\label{tab:sent_Values}
\end{table}

\begin{table}[h]
	\resizebox{\linewidth}{!}{\begin{tabular}{| p{3.0cm} | p{1.6cm} | p{1.2cm} |p{1.2cm} |p{1.3cm} |} 			\hline	
LIWC Category & Cohen's $d$ & Anon & Known & Diff. (\%) \\ \hline

Leisure	&	-0.273	&	0.009	&	0.018	&	-51.468 \\
Space	&	-0.252	&	0.052	&	0.068	&	-24.306 \\
Article	&	-0.217	&	0.062	&	0.076	&	-18.222 \\
Negemo	&	0.194	&	0.024	&	0.015	&	60.408 \\
Body	&	0.228	&	0.013	&	0.005	&	168.218 \\
SheHe	&	0.237	&	0.007	&	0.002	&	287.416 \\
Family	&	0.247	&	0.006	&	0.001	&	749.53 \\
Affect	&	0.275	&	0.068	&	0.047	&	43.305 \\
Verbs	&	0.283	&	0.161	&	0.14	&	15.161 \\
Friends	&	0.297	&	0.011	&	0.001	&	1224.947 \\
Present	&	0.347	&	0.129	&	0.104	&	24.684 \\
Sexual	&	0.357	&	0.013	&	0.001	&	1644.211 \\
Humans	&	0.359	&	0.019	&	0.005	&	310.755 \\
Funct	&	0.411	&	0.527	&	0.479	&	10.066 \\
Pronoun	&	0.504	&	0.13	&	0.086	&	50.945 \\
I	&	0.511	&	0.045	&	0.012	&	274.765 \\
Ppron	&	0.585	&	0.069	&	0.027	&	160.326 \\ \hline
EMPATH Category & Cohen's $d$ & Anon & Known & Diff. (\%) \\ \hline
Vehicle	&	-0.317	&	0.001	&	0.005	&	-84.005 \\
Air Travel	&	-0.269	&	0.001	&	0.004	&	-79.035 \\
Car		&	-0.261	&	0.001	&	0.004	&	-80.414 \\
Qar		&	-0.235	&	0.001	&	0.004	&	-73.046 \\
Magic	&	-0.234	&	0.000	&	0.002	&	-84.021 \\
Military	&	-0.226	&	0.001	&	0.004	&	-72.025 \\
Terrorism	&	-0.217	&	0.000	&	0.002	&	-88.607 \\
Technology	&	-0.199	&	0.008	&	0.015	&	-43.536 \\
Pain	&	0.195	&	0.006	&	0.002	&	257.665 \\
Masculine	&	0.197	&	0.004	&	0.001	&	495.7 \\
Affection	&	0.232	&	0.005	&	0.000	&	1356.032 \\
Childish	&	0.247	&	0.005	&	0.000	&	947.671 \\
Youth	&	0.269	&	0.006	&	0.001	&	779.78 \\
Positive Emotion	&	0.269	&	0.010	&	0.003	&	277.746 \\
Feminine	&	0.271	&	0.007	&	0.000	&	1234.727 \\
Love	&	0.272	&	0.008	&	0.001	&	533.962 \\
Wedding	&	0.308	&	0.010	&	0.002	&	531.320 \\
Children	&	0.308	&	0.010	&	0.002	&	473.889 \\
Sexual	&	0.314	&	0.010	&	0.001	&	1397.486 \\
Friends	&	0.366	&	0.021	&	0.006	&	249.363 \\ \hline
\end{tabular}}
\caption{Fraction of different lexical categories across the Anon and the Known clusters. Only cases with significant difference are shown.}
\label{tab:LIWC_Values}
\end{table}

\subsubsection{Sentiment analysis}
We perform sentiment analysis of the question text across clusters using VADER~\cite{hutto2014vader} to obtain the distribution of the positive, negative and neutral sentiments across the different clusters. We observe from Table~\ref{tab:sent_Values} that a major proportion of questions in both the Anon and the Known clusters have neutral sentiments (82.6\% and 87.4\% respectively). However, questions in the Anon clusters have relatively higher (75.28\%, Cohen's $d\sim0.25$) negative sentiments as compared to the Known clusters. These results are in correspondence with past observations that anonymous content generally bear more negative sentiments~\cite{correa2015many}.

\subsubsection{Psycholinguistic analysis}
In this section, we first perform LIWC analysis of the question text across the Known and the Anon clusters. We report the significantly different categories in Table~\ref{tab:LIWC_Values}. The categories which occur in significant higher proportions in the Anon clusters are intensely personal (\textit{`I', `Ppron} (\textit{Personal Pronouns})', \textit{`Body'}) and refer to relationships and social ties (\textit{`friends', `family', `sexual'}). Categories predominant in non-anonymous clusters (\textit{`leisure',  `space'}) refer to impersonal general topics like games, sports, party, travel and the like.
We also use Empath~\cite{fast2016empath} to lexically analyze text across more than 200 pre-trained categories. The questions in Known clusters have a higher proportion of words related to \textit{`vehicle air travel', `car', `war', `magic', `military', `terrorism' and `technology'}. On the other hand, the Anon clusters have more words related to \textit{`pain', `masculine', `affection', `childish', `youth', `positive emotion', `feminine', `love', `wedding', `children', `sexual' and `friends'}. These results are in agreement with the previous observations that the Anon clusters talk more about personal questions whereas the Known clusters talk more about general questions.

\subsubsection{Personality analysis}
In this section, we use the outcomes of the LIWC analysis in the previous section to identify personality traits of the users posting questions in those clusters. Since it is well-known that the personality of a person influences the dis-inhibition effect~\cite{suler2004online}, we build up on the research of Yarkoni~\shortcite{yarkoni2010personality} to find differences in personality traits. We use the Big-five personality traits method~\cite{costa92} and follow a method similar to that outlined in Bhattacharya and Ganguly~\shortcite{bhattacharya2016characterizing} to identify the personality. For readability we describe the Big-five method in brief below.

\noindent\textbf{Big-five personality traits}: The Big-five personality traits~\cite{costa92} is a mechanism to model human personality along five broad dimensions: `openness', `conscientiousness', `extraversion', `agreeableness', and `neuroticism'. Each of these individually refer to five different and non-overlapping personality traits related to human behavior. Here, we try to characterize the differences in the user personality when they post a question as anonymous compared to when they post a question as non-anonymous. Since, we don't have the information about the users (when they post as anonymous), we perform the personality analysis at the cluster level.

\noindent\textbf{Discovering personality traits}: We follow a method similar to that outlined in Bhattacharya and Ganguly~\shortcite{bhattacharya2016characterizing} to identify the traits. For both Anon and Known clusters, we first compute the mean percentage of words belonging to the different LIWC categories. As we want to study the personality traits of the Anon cluster, we select only those LIWC categories which maintain a significant positive difference between the Anon and Known group. The positive significance is determined by a Cohen's $d$ value of at least 0.2. Next, for each LIWC category we look for personality traits that are shown to have strong correlations in~\citet{yarkoni2010personality} and associate it either positively or negative.

For example, the LIWC categories `sexual', `body' and `friends' are known to positively correlate with the Big-five trait `extraversion'~\cite{yarkoni2010personality}. Now, if the Anon clusters maintain the significant Cohen's $d$ value with respect to the Known clusters for these three LIWC categories, then we would resolve that the users posting in the Anon clusters would exhibit \textit{more `extraversion'} in their behavior.

\noindent\textbf{Key insights}: The observations for the personality analysis are displayed in Table~\ref{tab:personality_quest_ratio}. A $+x$ $(-y)$ in the first row of the table indicates the number of significant LIWC categories that positively (negatively) correlate with the personality trait `extraversion'. Note that $x+y$ is the total number of LIWC categories that maintain a Cohen's $d$ value of $\geq 0.2$ between the Anon and the Known clusters. We see, that the Anon group has positive correlation with `extraversion' and`agreeableness'. In contrast, it seems to be negatively correlated to `openness'. People who have a higher score in `extraversion' tend to be more assertive, bold and enthusiastic. Likewise, those scoring high on `agreeableness' tend to be more trustworthy and straightforward. These traits are strongly correlated to the anonymous users, who are more expressive since they do not have the fear of being judged. In contrast, people who score high on `openness' tend to have a general appreciation for art, sports, adventure, unusual ideas, imagination, curiosity, and variety of experience, which are shown to be less pronounced in the Anon clusters.

However, the predictions for `conscientiousness' and `neuroticism' are not clear from the results in Table~\ref{tab:personality_quest_ratio} as they contain mixed signals.

\begin{table}[htbp]
	\footnotesize
	\resizebox{\linewidth}{!}{\begin{tabular}{| p{2.0cm}  p{0.7cm}  p{7cm} |} 
			\hline
			\textbf{Personality trait}&  &\textbf{LIWC categories} \\ \hline
			\multirow{2}{*}{Extraversion}&(+6)& Humans, Affect,  Family, Friends, Sexual, Body \\
										&(0)& \\
                                        
            \multirow{2}{*}{Agreeableness}&(+5)& Pronoun, Family, Friends, Sexual, Body \\
										&(0)& \\
                                        
            \multirow{2}{*}{Openness}&(0)&\\ 
										&(-6)&Humans, Pronoun, Affect, Family, I, Present \\
		
			\multirow{2}{*}{Conscientiousness}&(+0)&  \\
											  &(-1)& Humans\\

			\multirow{2}{*}{Neuroticism}&(+1)& I\\
										&(-1)& Friends\\

			\hline
	\end{tabular}}
	\caption{Personality traits of the users posting in the 
Anon predicted by the relative use of words from the different LIWC categories.}
	~\label{tab:personality_quest_ratio}
\end{table}

\subsection{Characterizing topics via response ratio difference}

So far, we have represented a topic solely by the proportion of anonymous questions, without accounting for the responses garnered by the questions. We acknowledge that the anonymous and non-anonymous questions that constitute a topic need not behave in a similar fashion in terms of number or speed of responses. 
This motivates us to further ``deep dive'', wherein we quantify the topics in terms of the responses received by the anonymous and non-anonymous questions in that topic. We estimate the difference in the \textit{response ratio} ($d_r$) for every cluster which is defined as follows --

\begin{center}
\large
$\frac{\text{\# Answers to Anon questions}}{\text{\# Total Anon questions}} - \frac{\text{\# Answers to Known questions}}{\text{\# Total Known questions}} $
\end{center}

\begin{figure}[htb]
	\centering
	\includegraphics[width=.475\textwidth]{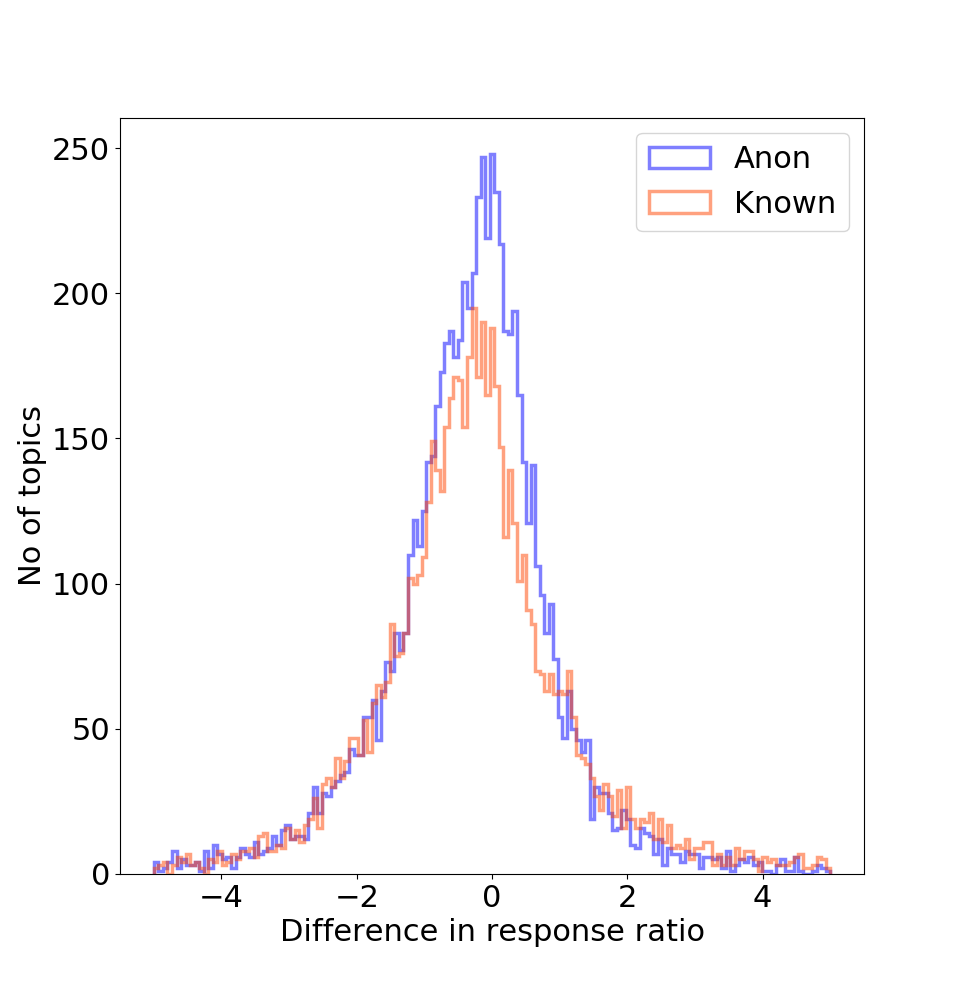} 
\caption{Response ratio differences across the different clusters.}
	~\label{fig:AR_Extreme}
    
\end{figure}

If the difference in response ratio ($d_r$) is positive (negative) then it indicates that the volume of answers corresponding to the anonymous questions within a topic is more (less) than the volume of answers corresponding to the non-anonymous questions. In 
Figure~\ref{fig:AR_Extreme}, we illustrate this difference in response ratio across the different Known and Anon clusters. 

\subsubsection{Definition of the extreme topics} Let the mean $d_r$ across all the topics be denoted by $\mu_{d_r}$; similarly, let the standard deviation be denoted by $\sigma_{d_r}$. We define a topic as \textit{positive} if its $d_r \geq \mu_{d_r} + \sigma_{d_r}$. Similarly, we call a topic as \textit{negative } if $d_r \leq \mu_{d_r} - \sigma_{d_r}$. Using this definition, we can have four groups based on the the topic belonging to the Anon or Known clusters. These four groups are :
\begin{itemize}
\item Anon Positive (\#clusters = 357) - \textit{Self-Medication, Borderline Personality Disorder, Burkha, Jealousy, Sexual Violence, Dealing with Loneliness, Adult Breastfeeding.}
\item Anon Negative (\#clusters = 435) - \textit{Dramatic Irony, Rednecks, Pet Advice, Bollywood Gossip, Wedding Night, Canadian Expatriates, Inside IIT, Ground Beef.}
\item Known Positive (\#clusters = 494) - \textit{Hypothetical Situation, Evolutionists, Collectivism and Individualism, Atheist Republic, China and Japan Tension, Bizarre Foods.}
\item Known Negative (\#clusters = 430) - \textit{Dialog Systems (software), Anachronisms, Competitor Research, Cricket Matches, Nobel Peace Prize Laureates, Life Experiences.}
\end{itemize}

The Anon Positive group (i.e., the Anon clusters wherein the volume of answers garnered by the anonymous questions is much higher than the volume of answers garnered by the non-anonymous questions ) presents highly sensitive and personal topics such as mental illness and depression, whereas the Known Negative group (i.e., the Known clusters wherein the volume of answers garnered by the non-anonymous questions is much higher than the volume of answers garnered by the anonymous questions) abounds in impersonal and general topics like technology. For our analysis, these two groups are the most important. Table~\ref{tab:question_pos_anon_neg_non_anon} shows some representative questions from these two extremes.

\begin{table*}[htbp]
	\resizebox{1.05\textwidth}{!}{\begin{tabular}{| p{10cm} | p{9.5cm} |} 
			\hline
			\textbf{Positive Anon} &\textbf{Negative Known} \\ \hline
			Has anyone with borderline personality disorder ever tried electroconvulsive therapy?&How does the "siri clone" cydia sara actually work, technically speaking?\\ 
			Why do people with bpd need to move so frequently? &Are there any chatbots based on irc logs?\\ 
			How did you find out you have borderline personality disorder?&What is the difference between wit.ai, api.ai, and askziggy?\\ 
			Can my borderline husband feel the depth of love that i feel&Can you train a content corpus in bluemix? \\ 
			How do you tell the differences between bpd, histronic, narcissism, antisocial, and bipolar?&What are examples of dialog systems based on the tasks of natural language processing\\ 
			How can people with bpd make rational decisions?&What's good and bad about chatbots"\\ 
			Are there degrees of borderline personality disorder like on a spectrum?&What are important and useful papers to read on dialogue systems?\\

			\hline
	\end{tabular}}
	\caption{Representative questions from the (i) Anon clusters with positive response ratio and (ii) Known clusters with negative response ratio.}
	~\label{tab:question_pos_anon_neg_non_anon}
\end{table*}

\subsubsection{Linguistic analysis of the extremes} 
We carry out the linguistic analysis, as performed in the previous section, on the topical extremes. The numbers in the parenthesis below indicate the significance in the differences in terms of Cohen's $d$. POS-TAG analysis shows that the differences in noun (-0.43), proper-noun (-0.22) and pronoun (0.54) categories have got amplified. We also find significant (0.233) increase in the proportion of negative sentiments in the positive Anon class. Some of the LIWC categories with amplified positive differences are `family' (0.32), `friends' (0.29),  `humans' (0.52), `I' (0.53), `Ppron' (0.56), `sexual' (0.37). Likewise, the EMPATH categories which have also been amplified are 'politics' (-0.212), 'fun' (-0.202), 'youth' (0.39), 'wedding' (0.44) and 'children' (0.42). The observations show that the linguistic differences are indeed more pronounced amongst these two groups. 

\section{Community perspective of anonymity}
\subsection{Anonymity of answers}
Thus far we have been designating a question as anonymous if the user posting the question would tag it as anonymous. This is therefore the user's perception about the question. In this discussion, we would like to point out that there could be a community perception about the same question and this might well be different from the individual's perception. One way to quantify the perception of the community could be to observe the \textit{level of anonymity in the answers} garnered by a question. Next we define this concept precisely.

For every question posted by a user, we compute the number of answers garnered by it. We note that it is possible that a question actually posted as anonymous can have a larger proportion of answers that are non-anonymous. The individual perception of the user asking the question therefore, in this case, is different from the community (i.e., users answering the question) perception. While the individual might keep himself/herself anonymous owing to some form of perceived stigma/taboo associated with the question, the community might not resonate with the same perception. Similarly, the other side of the coin is also equally possible, i.e., a question actually posted as non-anonymous by a user might garner a high proportion of anonymous answers. Based on this categorization, one can easily conceive of an \textit{anonymity grid} as follows. 

\subsection{Anonymity grid}

To better understand this, we combine the user and the community perspective of anonymity. The four groups so formed -- (i) (question Anon, answer Anon), (ii) (question Anon, answer Known), (iii) (question Known, answer Anon) and (iv) (question Known, answer Known) constitute a $2\times2$ table which we refer to as the anonymity grid. In Table~\ref{tab:anonymity_grid}, we report some of the representative Quora topics in each of these four classes. We observe that most of the taboo and sensitive topics correspond to the class (i) while the most general topics correspond to the class (iv).

\begin{table*}[h!]
	\tiny
	\resizebox{\linewidth}{!}{\begin{tabular}{| p{1cm} | p{4.5cm} | p{4.5cm} |} 
			\hline
			&\textbf{Question Anon} &\textbf{Question Known} \\ \hline
			\textbf{Answer Anon}&
			
			advice about love, adult dating advice, love and relationship advice, breaking up, romantic relationships, girlfriends, relationships and sex, dating advice for young people, dating and friendship, love life advice, marriage advice, dating and relationship advice, boyfriends, romance love, adult dating and relationships, dating advice, relationships and dating of young people, love, sexuality and relationships, relationship advice 
			
			&
			
			experiences in life, bharatiya janata party bjp, narendra modi, god, question that contains assumptions, culture of india, life and living, politics of india, humor, what does it feel like to x, advice for everyday life, muslims, bollywood, people, philosophy of everyday life, israel, islam, indian ethnicity and people, spirituality, hinduism\\ 
            \hline
			
			\textbf{Answer Known}&
			
			dog breeds, obstetrics and gynecology, acne, womens health, menstruation, programming interviews, hairstyles, dermatology, english sentences, purpose, phrase definitions, pregnancy, skin care, teeth, pets, google recruiting, phrases, emotional advice, makeup cosmetics, hair care
			
			&
			
			major league baseball, international space station, seo tools, black holes, solar panels, aircraft, military technology, general relativity, camera lenses, augmented reality, ipad applications, spacetime, virtual reality, military aircraft, solar energy, nuclear energy, hypothetical planetary science scenarios, satellites, google analytics product, national football league nfl\\

			\hline
	\end{tabular}}
	\caption{Representative topics in the anonymity grid.}
	~\label{tab:anonymity_grid}
\end{table*}
In order to find topics which belong to the different groups in the anonymity grid, we first filter out all the topics which have less than 1000 questions. Then, for each topic we find the fraction of questions in each of the cell in the anonymity grid. We then simply rank the topics in each cell based on this fraction.  Table~\ref{tab:anonymity_grid} shows the top 20 topics in each cell of the grid. The topics obtained as a result of the ranking algorithm resonate well with our expectations.
\begin{itemize}
\item The category (question Anon, answer Anon) has a large proportion of topics pertaining to relationships, sex, dating etc. which are highly sensitive and thus justifiably anonymous in terms of both questions and answers.
\item The category (question Anon, answer Known) contains topics related to health, jobs, careers, personal grooming, pets etc. It has questions which are highly personal to the user (e.g., `How do i get rid of acne?') who wish to seek advice anonymously. The thriving userbase of Quora which has many domain experts and the desire for social recognition drives the answers in this category to be predominantly non-anonymous.
\item The category (question Known, answer Anon) contains topics like politics, religion and experience/advice. Answers in such topics might often lead to controversy and could generate backlash for the user. Thus, these questions are often answered anonymously although the question itself is posted as non-anonymous, since the main target of the backlash and controversy are the answerers and not the user posting the question.
\item The category (question Known, answer Known) contains very general topics like sports, science and technology etc. where both the answers and questions are posted non-anonymously.
\end{itemize}

\section{First Response Time}
In addition to observing the number of responses garnered by a question, it is also interesting to note the \textit{first response time} (henceforth FRT). We define FRT of a question as the amount of time elapsed (in hours) between posting the question and receiving the first response. We can then compare the FRT of the anonymous questions and non-anonymous questions within a topic by observing their corresponding median FRT. We consider the median FRT to prevent the skewness arising from unanswered questions or questions answered after a long period of time (1-2 years).

\subsection{FRT for the global and topical scale}

We present a comparison of the FRT for the anonymous and non-anonymous questions on the global, topical and extremes level in Table~\ref{tab:topics-response-time}. We observe that on a global scale, there is not much difference in the median FRT for the Anon and the Known questions. The same observation holds true for the Anon and Known questions 
within the individual Anon and Known Clusters. We thus, note that on a global or topical scale, the nature of the question is not instrumental to the question's FRT. However, it is important to note that the median FRT for all the questions in the Known clusters is 66\% more than that of the Anon clusters. This illustrates that topics which are predominately anonymous receive a faster response.
The topical extremes (as defined in the previous section), also demonstrate stark differences. It is observed that the response time is lower for the Anon questions in both Anon Positive and Known Positive categories as opposed to the Known questions in those categories. Likewise, the response time is lower for the Known questions in the Anon Negative and Known Negative categories as opposed to the Anon questions.

\begin{table}[htbp]
	{
    \footnotesize
    \centering
    \begin{tabular}{|p{22mm}|p{14mm}|p{14mm}|p{14mm}|}
	\hline
	Category& Anon Questions & Known Questions & Diff (in $\%$)\\ \hline
    Global & 9.317&9.117&2.194\\
    Anon Clusters & 4.883& 5.433& -10.123\\
    Known Clusters & 8.517& 8.55& -0.39\\
    Anon Positive & 3.342&	5.483&	-39.058\\
    Anon Negative& 12.8&	5.65&	126.549\\	
    Known Negative &8.1&	6.133&	32.065\\	
	Known Positive& 6.575&	10.55&	-37.678\\
    \hline
	\end{tabular}}
	\caption{Comparing the median first response time (FRT) of anonymous questions and non-anonymous questions on a global scale, Anon and Known topics and for the topical extremes.}
	~\label{tab:topics-response-time}
\end{table}

\subsection{FRT for the anonymity grid}
We consider the top 50 representative topics of each cell in the anonymity grid (described in earlier section). We then observe the differences in FRT of the anonymous and non-anonymous questions (of the representative topics) for each cell, as shown in Table~\ref{tab:anon-grid-response-time}. The categories in Table~\ref{tab:anon-grid-response-time} correspond to Anon question and Anon answer (Anon-Anon), Anon question and Known answer (Anon-Known), Known question and Anon answer (Known-Anon) and Known question and Known answer (Known-Known). 

It is interesting to note that the rows (1,3) which have a comparatively higher proportion of anonymous answers have a lower median FRT as compared to rows (2,4) which have a higher proportion of non-anonymous answers. This hints that the community's perspective of anonymity could be more instrumental in determining the response time of a question than an individual's. 

\subsection{FRT and community support}
A pertinent observation from the Tables~\ref{tab:topics-response-time} and~\ref{tab:anon-grid-response-time} is that the lowest median FRT is observed for the Anon Positive (3.34 hours) and Anon-Anon group (1.28 hours). The topics in these groups are extremely sensitive and deal with personal issues like `depression', `relationship advice', `self-medication', `love' etc. The extreme low values of FRT for the anonymous questions in these topics is indicative of the community's support. It also encourages users to post anonymously without the fear of backlash and judgement and fosters user engagement.

\begin{table}[htbp]
	{
    \footnotesize
    \centering
    \begin{tabular}{|p{22mm}|p{14mm}|p{14mm}|p{14mm}|}
	\hline
	Category& Anon questions & Known questions & Diff (in $\%$)\\ \hline
    Anon-Anon & 1.283 & 2.083 & -38.4 \\
    Anon-Known& 9.533 & 7.85 & 21.444 \\
    Known-Anon& 2.7& 3.2 & -15.625\\
    Known-Known&4.883 & 4.017 & 21.577 \\
    \hline
	\end{tabular}}
	\caption{Comparing FRT of representative topics of each cell in the anonymity grid. }
	~\label{tab:anon-grid-response-time}
\end{table}

\section{Conclusions and future works}

In this paper, we performed a large scale anonymity analysis on the popular Q\&A site Quora and found that on a global scale, no significant differences  exists between the anonymous and non-anonymous questions. However, nuanced linguistic and psycholinguistic differences become prominent when we topically cluster the questions. We observed that clusters with higher proportion of anonymous questions, have relatively higher negative sentiments and express more personal and sensitive information as opposed to those with higher proportion of non-anonymous questions. We also introduced the idea of an anonymity grid and studied how the perception of anonymity of the user posting the question differs from the community that answers it. We further introduced the concept of First Response Time and observed that clusters with high fraction of anonymous questions get faster response. In fact the fastest response is observed for the extremely personal topics related to anxiety, depression, sexual violence and is thus indicative of the community's support, thereby, fostering user engagement.

As a part of future work, we would investigate the temporal aspects of anonymity. Specifically, we would like to ask what kind of changes does a topic undergo from the perspective of anonymity. There would be some topics which would have been predominantly anonymous initially, but as the society progresses and social norms change, people might start to post questions/answers non-anonymously in that topic causing a gradual shift in the anonymity ratio of the topic. Another interesting direction would be to analyze the anonymity from a demographic and gender perspective using APIs such as NamePrism\footnote{http://www.name-prism.com/} and Genderize.io\footnote{https://genderize.io/} respectively.

\bibliography{Main}
\bibliographystyle{aaai}

\end{document}